\documentclass[cits]{PoS}
\usepackage{amsmath}
\newcommand{\FC}{\;,}
\newcommand{\FD}{\;.}
\newcommand{\I}{\mathrm{i}}  
\newcommand{\E}{\mathrm{e}}  
\newcommand{\myparagraph}[1]{~\vspace*{-2mm}\\ \noindent{\bf #1}}

\title{$K\pi$ scattering in moving frames}

\ShortTitle{$K\pi$ scattering in moving frames}

\author{\speaker{C. B. Lang}\\
        Institut f\"ur Physik,  University of Graz, A--8010 Graz, Austria\\
        E-mail: \email{christian.lang@uni-graz.at}}
\author{Luka Leskovec\\
        Jozef Stefan Institute, 1000 Ljubljana, Slovenian\\
        E-mail: \email{luka.leskovec@ijs.si}}
\author{Daniel Mohler\\
        Fermi National Accelerator Laboratory, Batavia, Illinois  60510-5011, USA\\
        E-mail: \email{dmohler@fnal.gov}}
\author{Sasa Prelovsek\\
        Department of Physics, University of Ljubljana and Jozef Stefan Institute, 1000 Ljubljana, Slovenia\\
        E-mail: \email{sasa.prelovsek@ijs.si}}

\abstract{We extend our study of the $K\pi$ system to moving frames and present an exploratory
extraction of the  masses and widths for the $K^*$ resonances by simulating $K\pi$ 
scattering in $p$-wave with $I=1/2$ on the lattice. Using $K\pi$ systems with 
non-vanishing total momenta allows the extraction of  phase shifts at several values 
of $K\pi$ relative momenta. A Breit-Wigner fit of the phase renders a $K^*(892)$ 
resonance mass and $K^*\to K \pi $ coupling compatible with the experimental numbers.
We also determine the $K^*(1410)$ mass assuming the experimental $K^*(1410)$ width. 
We contrast the resonant $I=1/2$ channel with the repulsive non-resonant 
$I=3/2$ channel, where the phase is found to be negative and small, in agreement with 
experiment.  
}

\FullConference{31st International Symposium on Lattice Field Theory - LATTICE 2013\\
		July 29 - August 3, 2013\\
		Mainz, Germany}

\begin{document}

\section{Overview}

In the lattice studies of meson excitations is has become clear that it is not sufficient to use 
quark-antiquark interpolators only. Although the dynamical quark vacuum allows in principle the
coupling of $\bar q q$ operators to meson-meson intermediate states, those have usually not been
observed in the correlation functions for  $\bar q q\to \bar q q$  alone. If one enlarges the set
of interpolators to include also 4-quark (meson-meson) operators the full system does exhibit
energy levels related to these states.

In \cite{Lang:2012sv} we have studied the $K\pi$ system for both isospin values $\frac{1}{2}$ and $\frac{3}{2}$
in $s$-wave and $p$-wave in the rest frame. Due to the small lattice size $m_\pi L\simeq 2.7$ only one energy
level was in the resonance region of the $K^*(892)$ vector meson. Changing to interpolators in
moving frames allows us to obtain further energy levels on the same set of configurations. For that
reason we extended our analysis for the $p$-wave to values of non-zero total momentum.
The determination of $K^*(892)$ is challenging even in this case since the experimental width is only 50 MeV and 
even smaller for $m_\pi>m_\pi\textrm{phys}$.

Scattering of two mesons is described by correlation functions $C_{ij}(t)=\langle \mathcal{O}_i(t)\mathcal{O}
_j(0)\rangle$ where $t$ denotes the Euclidean time and $\mathcal{O}$ operators with the channel's quantum numbers. The 
energy spectrum obtained from the eigenvalues of such propagators for finite spatial volumes is discrete. Under 
certain conditions (localised interaction region smaller than the size of the volume) one may relate these energy 
levels to values of the scattering phase shift in the continuum. The corresponding relations were first formulated for 
two identical particles in the rest frame by L{\"u}scher  \cite{Luscher:1990ux,Luscher:1991cf}, for moving frames in 
\cite{Rummukainen:1995vs,Kim:2005zzb} and for mesons with unequal masses in
 \cite{Fu:2011xz,Leskovec:2012gb,Gockeler:2012yj}. Moving frames considerably complicate the analysis. Unfortunately, for moving frames 
there occurs mixing of partial waves and  due to
relativistic distortion the symmetry group changes from the simple cubic $O_h$ to dihedral subgroups. One has to 
write the interpolators as representations of such groups.

Here we briefly discuss the main steps of our analysis and the principal results. More details can be found in \cite{Lang:2012sv,Prelovsek:2013ela}.

\section{Analysis Details}

\myparagraph{Configurations: }
The analysis is based on one ensemble of gauge configurations on lattices of size $16^3\times 32$, with clover Wilson dynamical, mass-degenerate $u,\,d$ quarks and 
$u,\,d,\,s$ valence quarks ($m_s>m_u=m_d$).  This ensemble has been generated by the authors of  \cite{Hasenfratz:2008ce,Hasenfratz:2008fg} while studying re-weighting techniques.  The corresponding pion mass is $m_\pi= 266(4)~$MeV. The strange quark valence mass is fixed by $m_\phi$, which corresponds to  $\kappa_s=0.12610$ and $c_{sw}=1$, giving $m_K= 552(6)~$MeV. The parameters of the ensemble are given in \cite{Lang:2011mn,Lang:2012sv,Prelovsek:2013ela}. Due to the limited data on just a single ensemble, our determination of the lattice spacing $a=0.1239(13)$ fm results from taking a typical value of the Sommer parameter $r_0=0.48$ fm. We note that the uncertainty associated with this choice might lead to small shifts of all dimensionful quantities.  

\myparagraph{Variational analysis: }
We determine the energy levels of the coupled $K^*$
and  $K\,\pi$ system with help of the variational 
method \cite{Michael:1985ne,Luscher:1985dn,Luscher:1990ck,Blossier:2009kd}. For a given
quantum channel one measures the Euclidean cross-correlation matrix $C(t)$
between several interpolators living on the corresponding Euclidean time slices. The
generalized eigenvalue problem  disentangles the
eigenstates $|n\rangle$. From the exponential decay of the
eigenvalues one determines the energy values of the eigenstates by exponential fits to the
asymptotic behavior. 
In order to reliably obtain the lowest energy eigenstates and energy levels one needs a sufficiently 
large set of interpolators with the chosen quantum numbers. 

\myparagraph{Wick contractions: }  
In order to compute this correlation matrix one has to compute the Wick decomposition of the correlators in terms of the quark propagators. There are no completely disconnected terms, however there are contributions with backtracking quark lines. This necessitates an efficient algorithm to reliably determine those.

\myparagraph{Distillation: } 
We used the so-called distillation method \cite{Peardon:2009gh}. On a given time slice one
introduces separable quark smearing sources derived from the
eigenvectors of the spatial lattice Laplacian. This allows high flexibility due to the disentanglement of the computation of the quark propagators (``perambulators'') and the hadron operators.

\myparagraph{Phase shifts: } 
There are relations between the energy levels at finite volume and the phase shifts of the infinite volume, valid in the elastic region \cite{Luscher:1990ux,Luscher:1991cf,Rummukainen:1995vs,Kim:2005zzb,Leskovec:2012gb,Gockeler:2012yj}.
We apply  these up to the region of the second resonance, as discussed below. More detailed studies would have to include more operators and deal with the coupled channel problem. 

\myparagraph {Moving frames: } 
In order to facilitate the $K^*\to K\pi$ decay kinematically and to access further values of the center-of-momentum frame (CMF) energy, we implement interpolators 
with vanishing as well as  non-zero total momenta $\vec{P}$ (we considered also all  permutations of $\vec{P}$ and all possible directions of polarizations at given $\vec{P}$). The relativistic distortion of the lattices reduces the symmetries: 
\vspace*{-2mm}
\begin{align}\label{irreps}
\vec{P}&\!=\!\tfrac{2\pi}{L}\vec{e}_z:\,           &C_{4v}\,, ~&\mathrm{irreps}\;E(\vec{e}_{x,y}),~E(\vec{e}_x\pm \vec{e}_y),~ l=1,2\FC\nonumber \\
\vec{P}&\!=\!\tfrac{2\pi}{L}(\vec{e}_x+\vec{e}_y): &C_{2v}\,, ~&\mathrm{irreps}\;B_2,~ B_3,\qquad\qquad~\quad l=1,2\FC\nonumber\\
\vec{P}&\!=\!0:                         &O_h\,, ~     &\mathrm{irrep}\;T_1^-,\qquad\qquad\qquad\quad~  l=1\FD
\end{align}
The zero-momentum case, studied in \cite{Lang:2012sv}, is listed for completeness since we combine all these results.   The analytic framework for  $p$-wave scattering using the first two  momenta is described in detail in \cite{Leskovec:2012gb}, together with the symmetry considerations, appropriate interpolating fields and extraction of the phase shifts. 
 
The symmetry group of the  mesh viewed from the CMF of the $K\pi$ system is $C_{4v}$ for $\vec{P}=\tfrac{2\pi}{L}\vec{e}_z$ and $C_{2v}$ for $P=\tfrac{2\pi}{L}(\vec{e}_x+\vec{e}_y)$. These groups do not contain the inversion as an element, which  in turn implies that even and odd partial waves can in principle mix within the same irreducible representation. 
An example would be the irrep $A_1$, where $\delta_{l=1}$ mixes with $\delta_{l=0}$, since  $\delta_{0}(s)$ is known to be non-negligible in the whole energy region of interest.  However, $\delta_1$ does not mix with  $\delta_0$ in the irreducible representations $E,~B_2,~B_3$ (see equation (\ref{irreps})), so we can use these. In fact we employ two  representations of the two-dimensional $E$: $E(\vec{e}_{x,y})$ with basis vectors along axis $(\vec{e}_x,\vec{e}_y)$ and $E(\vec{e}_x\pm \vec{e}_y)$ with basis vectors along the diagonal $(\vec{e}_x+\vec{e}_y,\vec{e}_x-\vec{e}_y)$. 

\myparagraph{Interpolators: }
The interpolators are constructed to transform according to irreducible representations $B_2$, $B_3$, $E(\vec{e}_{x,y})$, $E(\vec{e}_x\pm \vec{e}_y)$, and $T_1^-$. This will give energy levels $E_n$, values of $s=E_n^2-\vec{P}^2$ and scattering phases $\delta(s)$.  

We use five operators (for $I=1/2$) combining terms of the type
\begin{equation}
\mathcal{O}(\vec{p},t)=\sum_{\vec{x}}\; \E^{\I\vec{p}\cdot\vec{x}}\,\overline s(\vec{x},t)\, \Gamma\, u(\vec{x},t) \;,
\end{equation}
where $\Gamma$ includes Dirac structure and derivatives. The combinations depend on the
used representations of the symmetry groups. For $I=3/2$ there are no quark-antiquark interpolators.

In addition to the quark-antiquark operators we used two $K\pi$ interpolators; these are linear combinations of $K(\vec{p}_K)\pi(\vec{p}_\pi)$ where momenta for $K$ and $\pi$ are separately projected  
\begin{equation}
K^+(\vec{p}_K,t)=\sum_\mathbf{\vec{x}}\;\E^{\I \vec{p}_K\cdot \vec{x}}\,\bar s(\vec{x,t})\, \gamma_5\, u(\vec{x,t})\FC \quad
\pi^+(\vec{p}_\pi,t)=\sum_{\vec{x}}\;\E^{\I \vec{p}_\pi \cdot \vec{x}}\,\bar d(\vec{x},t)\, \gamma_5\, u(\vec{x,t})\FC
\end{equation}
and analogously for $K^0$ and $\pi^0$. 
We simulated three permutations of direction $P=\tfrac{2\pi}{L}\vec{e}_z$, and  three  of direction $P=\tfrac{2\pi}{L}(\vec{e}_x+\vec{e}_y)$.
The complete list is given in \cite{Prelovsek:2013ela}.

Experience  shows that in the elastic region one typically has only three situations:
\begin{itemize}
\item Two-meson levels close to the energy of the non-interacting two meson system; the energy shift is small and the corresponding phase shift close to multiples of $\pi$.
\item Isolated resonance levels with little influence from the two-meson states; that is the case, when the resonance width is much smaller that the next nearby two-meson level.
\item Energy levels in a region where two-meson states and the resonance state superimpose, the so-called region of avoid level crossing.
\end{itemize}
In fact, by including or removing the two-meson interpolators with different relative momenta one usually sees quite clearly an added or removed energy level and whether the other energy levels are shifted. The picture emerges that the quark-antiquark interpolators couple (and generate) the generic resonance states and the two-meson operators the generic two-meson states and only in the region, where their
respective eigen-energies overlap, there are considerable effects on energy levels already 
observed in either the quark-antiquark or meson-meson basis. This is essentially the unitarization effect.

One thus observes, that the two-meson interpolator with higher relative momenta have little influence on the low lying eigenstates. Since (at least in the elastic region) there are -- depending on the spatial size $L\,m_\pi$ -- only a few (in our case one or two) such operators,
we are confident, that the observed levels would not change when adding further interpolators. 

\begin{figure}[tbp]
\begin{center}
\includegraphics*[width=0.75\textwidth,clip]{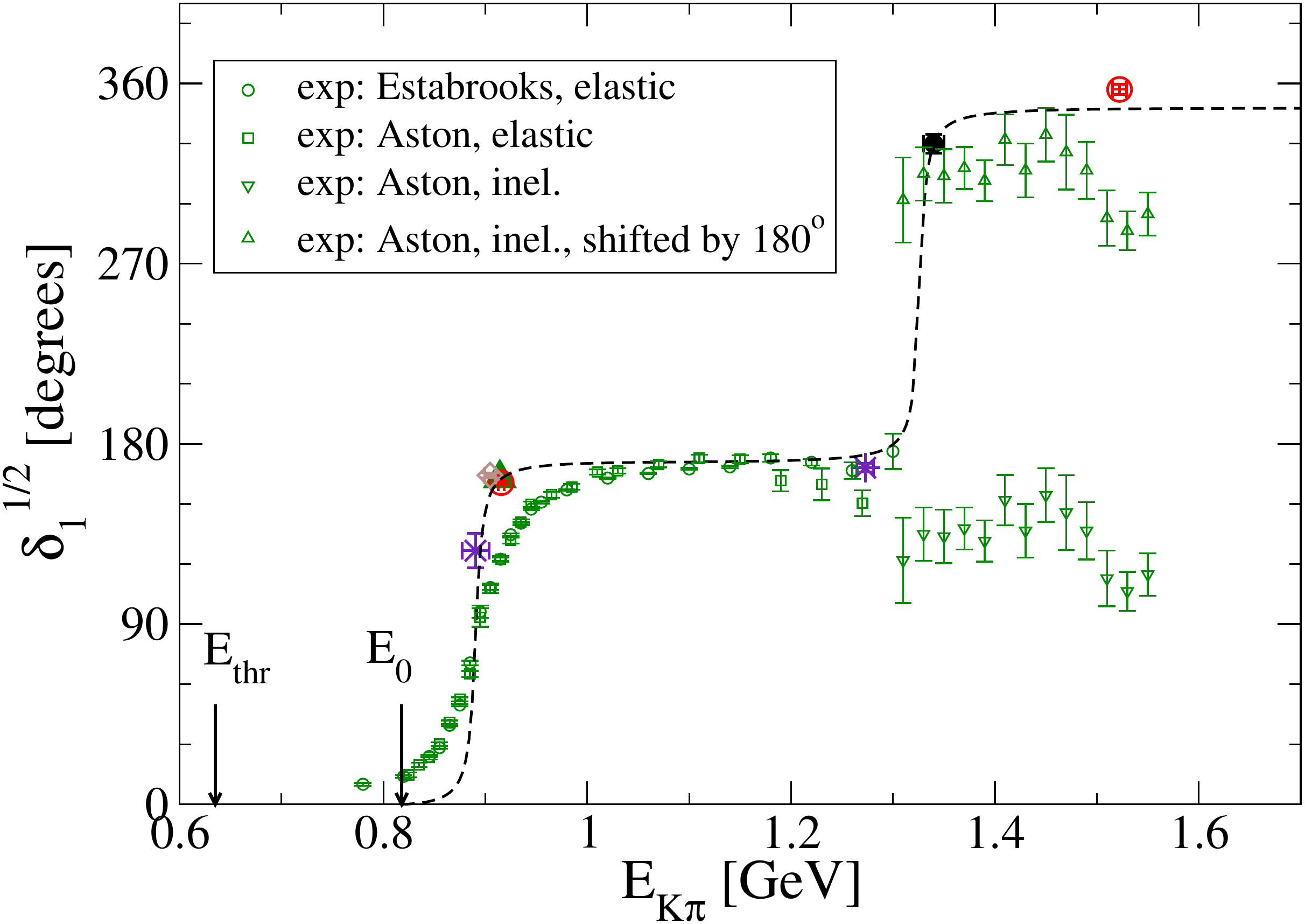}
 \end{center}
\caption{The $p$-wave scattering phase shift $\delta_{l=1}$, $I=1/2$  as a function of $E_{K\pi}\equiv\sqrt{s}$.  Different colors/symbols indicate results from different irreducible representations, while the point at $E_{K\pi}\simeq 1.34$ GeV, obtained by taking into account  $\delta_{1,2}$ mixing, is indicated by black dot. Note that three points (circle, triangle and diamond) near $\sqrt{s}\simeq 0.91~$GeV are overlapping. The  dashed line represents a fit over a pair of Breit-Wigner resonances as discussed in the text. The arrows indicate the physical threshold $E_{thr}$ and the location
$E_0=m_K+m_\pi$ of the threshold for our unphysical (larger) masses. We also show experimental results  (green symbols) by Estabrooks et al. \cite{Estabrooks:1977xe} and Aston et al. \cite{Aston:1987ir}. Due to the definition of the phase shift modulo $\pi$ we also plot the phase shift in the inelastic region  \cite{Aston:1987ir} incremented by $\pi$.}
\label{fig:delta12}
\end{figure}  
\begin{figure}[tbp]
\begin{center}
\includegraphics*[width=0.75\textwidth,clip]{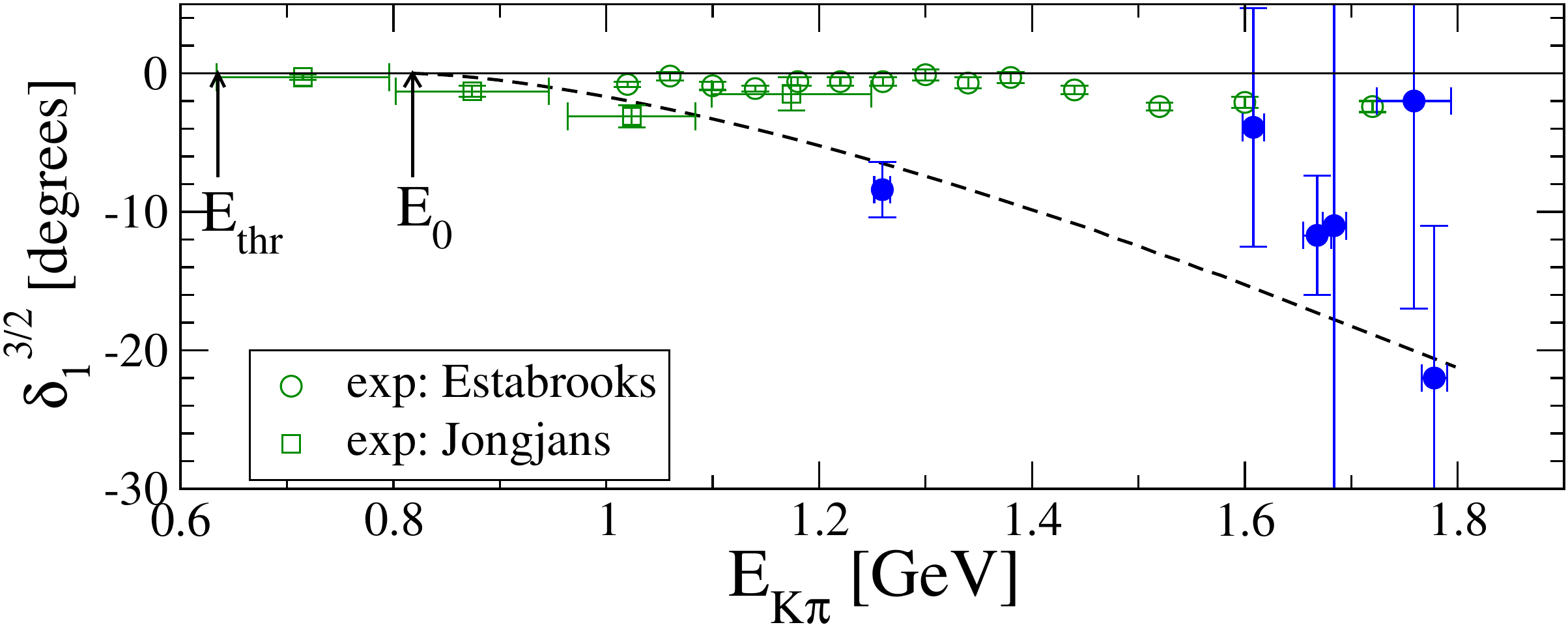}
\end{center}
\caption{ The $p$-wave scattering phase shift $\delta_{l=1}$, $I=3/2$. as a function of $E_{K\pi}\equiv\sqrt{s}$. Our results are given by blue circles, the green symbols denote experimental values due to \cite{Estabrooks:1977xe} and \cite{Jongejans:1973pn}. The broken line indicates a one-parameter fit (effective range fit with vanishing range parameter) to our data, setting $(p^3/\sqrt{s})\cot\delta(p)=\alpha$. The arrows indicate the physical threshold $E_{thr}$ and the location
$E_0=m_K+m_\pi$ of the threshold for our unphysical (larger) masses. }
\label{fig:delta32}
\end{figure}  

\section{Results}

The phase shift for $K\pi$ scattering was extracted from experiment long time ago \cite{Jongejans:1973pn,Estabrooks:1977xe,Aston:1987ir}. 
In phenomenological studies the $K^*$ resonance-pole emerged for example within the Roy-equation  approach and/or unitarized versions of the Chiral Perturbation Theory (for references see \cite{Prelovsek:2013ela}.

In \cite{Prelovsek:2013ela} we present a table of the energy levels resulting from our analysis. The corresponding phase shift values are compared with the experimental values in Figs. \ref{fig:delta12} and  \ref{fig:delta32}.

\myparagraph{The  $I=1/2$ channel: }
We obtain four phase shift points in the vicinity of $K^*(892)$ indicating a fast rise of the phase in a narrow region around $E_{K\pi}\equiv\sqrt{s}\simeq 0.89\;{\mathrm GeV}\simeq m_{K^*}$. Note that phase shift points from $B_3$, $E$ and $T_1^-$, that  almost overlap in $E_{K\pi}$, overlap also in $\delta_1$; this is non-trivial since L\"uscher's relations for these three irreducible representations have a different form.

Due to our unphysical values of the pion and the kaon masses, the threshold is closer to the resonance position than in nature, leading to a smaller width. A Breit-Wigner fit gives a mass of $m_{K^*(892)}=0.891(14)$ GeV and  a coupling $g_{K*K\pi}=5.7(1.6)$ (to be compared with the value 5.72(6) from experiments).

Above $E_{K\pi}\simeq1.3$ GeV the process becomes inelastic; for this exploratory study we still assume elasticity and apply the L\"uscher-tpye relations. 
Another complication arises from the fact that $d$-wave phase shift $\delta_2$ cannot be neglected around $\sqrt{s}\simeq m_{K_2^*(1430)}$. We therefore derived L\"uscher relations that contain $\delta_1$ as well as $\delta_2$ for irreps considered here: they are obtained from the so-called  determinant condition. Combining results from different representation allowed us to disentangle $p$- and $d$-wave (for details see \cite{Prelovsek:2013ela}).

We attempt an  extraction of the $K^*(1410)$ resonance parameters by fitting the resulting $\delta_1$ using a Breit-Wigner parametrization for two  resonances in the elastic region,
\begin{equation}\label{xy_sum}
\frac{p^{*3}}{\sqrt{s}}\cot\delta_1(s)=\biggl[\sum_{K_i^*}\;\frac{g_{K_i^*}^2}{6\pi}\;\frac{1}{m_{K_i^*}^2-s}\biggr]^{-1}\FC\quad
K_i^*=K^*(892),~K^*(1410)\FD\end{equation}
We fix   $m_{K^*(892)}$ and $g_{K^*(892)}$ in to the values  obtained from the single resonance fit in the $K^*(892)$ region, fixing
$g_{K^*(1410)}$ to the value $g_{K^*(1410)}^{exp}=1.59(3)$  derived from $\Gamma^{exp}[K^*(1410)\to K\pi]$ and get
a resonance position of $m_{K^*(1410)}= 1.33(2)$ GeV.

\myparagraph{The  $I=3/2$ channel: }
We extract $\delta_1^{3/2}$ assuming elasticity  and $\delta_2^{3/2}=0$, employing the same phase shift relations as for $I=1/2$.
The resulting phase shift in  Fig. \ref{fig:delta32} is  small and negative (consistent with zero). We also show the result of a one-parameter
fit to the leading term of an effective range approximation, $(p^3/\sqrt{s})\cot\delta(p)=\alpha$.
 
At this conference the Hadron Spectrum Collaboration \cite{Wilson:2013xxx} presented results for $I=3/2$ $K\pi$ phase shift \cite{Wilson:2013xxx} derived using techniques similar to ours; whereas our results from \cite{Lang:2012sv} agrees for the $s$-wave we disagree in the sign for the $p$-wave. Although we have different quark masses (in \cite{Wilson:2013xxx} the pion mass is
larger) this discrepancy is considerable and should be clarified.

\myparagraph{Acknowledgments: }We  thank Anna Hasenfratz for providing the
gauge configurations used for this work.  We would like to thank J.~Bulava, S.~Descotes-Genon, C.~Morningstar and  C.~Thomas for valuable discussions. The calculations were performed at Jozef Stefan Institute. This work is supported by the Slovenian Research Agency. Fermilab is operated by Fermi Research Alliance, LLC under Contract No. De-AC02-07CH11359 with the United States Department of Energy.

\providecommand{\href}[2]{#2}\begingroup\raggedright\endgroup


\begin{thebibliography}{10}

\bibitem{Lang:2012sv}
C.~B. Lang, L.~Leskovec, D.~Mohler, and S.~Prelovsek, {\it {$K \pi$ scattering for
  isospin 1/2 and 3/2 in lattice QCD}},  {\em Phys. Rev. D} {\bf 86} (2012)
  054508, [\href{http://arxiv.org/abs/1207.3204}{{\tt arXiv:1207.3204}}].

\bibitem{Luscher:1990ux}
M.~L{\"u}scher, {\it {T}wo-{P}article states on a torus and their
  relation to the scattering matrix},  {\em Nucl. Phys. B} {\bf 354}
  (1991) 531.

\bibitem{Luscher:1991cf}
M.~L{\"u}scher, {\it {S}ignatures of unstable particles in finite volume},
  {\em Nucl. Phys. B} {\bf 364} (1991) 237.

\bibitem{Rummukainen:1995vs}
K.~Rummukainen and S.~Gottlieb, {\it {R}esonance scattering phase shifts
  on a non-rest frame lattice},  {\em Nucl. Phys. B} {\bf 450} (1995)
  397, [\href{http://arxiv.org/abs/hep-lat/9503028}{{\tt hep-lat/9503028}}].

\bibitem{Kim:2005zzb}
C.~Kim, C.~T. Sachrajda, and S.~R. Sharpe, {\it {F}inite-volume effects in
  moving frames},  {\em Nucl. Phys. B} {\bf 727} (2005) 218,
  [\href{http://arxiv.org/abs/hep-lat/0510022}{{\tt hep-lat/0510022}}].

\bibitem{Fu:2011xz}
Z.~Fu, {\it {Rummukainen-Gottlieb's formula on two-particle system with
  different mass}},  {\em Phys. Rev. D} {\bf 85} (2012) 014506,
  [\href{http://arxiv.org/abs/1110.0319}{{\tt arXiv:1110.0319}}].

\bibitem{Leskovec:2012gb}
L.~Leskovec and S.~Prelovsek, {\it {Scattering phase shifts for two particles
  of different mass and non-zero total momentum in lattice QCD}},  {\em Phys.
  Rev. D} {\bf 85} (2012) 114507, [\href{http://arxiv.org/abs/1202.2145}{{\tt
  arXiv:1202.2145}}].

\bibitem{Gockeler:2012yj}
M.~G{\"o}ckeler, R.~Horsley, M.~Lage, U.~G. Mei{\ss}ner, P.~E.~L. Rakow,
  A.~Rusetsky, G.~Schierholz, and J.~M. Zanotti, {\it {Scattering phases for
  meson and baryon resonances on general moving-frame lattices}},  {\em Phys.
  Rev. D} {\bf 86} (2012) 094513,
  [\href{http://arxiv.org/abs/1206.4141}{{\tt arXiv:1206.4141}}].

\bibitem{Prelovsek:2013ela}
S.~Prelovsek, L.~Leskovec, C.~B. Lang, and D.~Mohler, {\it {$K \pi$ scattering and
  the K* decay width from lattice QCD}},  {\em Phys. Rev. D} {\bf 88} (2013)
  054508, [\href{http://arxiv.org/abs/1307.0736}{{\tt arXiv:1307.0736}}].

\bibitem{Hasenfratz:2008ce}
A.~Hasenfratz, R.~Hoffmann, and S.~Schaefer, {\it {Low energy chiral constants
  from epsilon-regime simulations with improved Wilson fermions}},  {\em Phys.
  Rev. D} {\bf 78} (2008) 054511, [\href{http://arxiv.org/abs/0806.4586}{{\tt
  arXiv:0806.4586}}].

\bibitem{Hasenfratz:2008fg}
A.~Hasenfratz, R.~Hoffmann, and S.~Schaefer, {\it {R}eweighting towards the
  chiral limit},  {\em Phys. Rev. D} {\bf 78} (2008) 014515,
  [\href{http://arxiv.org/abs/0805.2369}{{\tt arXiv:0805.2369}}].

\bibitem{Lang:2011mn}
C.~B. Lang, D.~Mohler, S.~Prelovsek, and M.~Vidmar, {\it {Coupled channel
  analysis of the $\rho$ meson decay in lattice QCD}},  {\em Phys. Rev. D} {\bf
  84} (2011) 054503, [\href{http://arxiv.org/abs/1105.5636}{{\tt
  arXiv:1105.5636}}].

\bibitem{Michael:1985ne}
C.~Michael, {\it {A}djoint sources in lattice gauge theory},  {\em
  Nucl. Phys. B} {\bf 259} (1985) 58.

\bibitem{Luscher:1985dn}
M.~L{\"u}scher, {\it {V}olume dependence of the energy spectrum in massive
  quantum field theories. {I}. {S}table particle states},  {\em Commun. Math.
  Phys.} {\bf 104} (1986) 177.

\bibitem{Luscher:1990ck}
M.~L{\"u}scher and U.~Wolff, {\it {H}ow to calculate the {E}lastic scattering
  matrix in 2-dimensional quantum field theories by numerical
  simulation},  {\em Nucl. Phys. B} {\bf 339} (1990) 222.

\bibitem{Blossier:2009kd}
B.~Blossier, M.~DellaMorte, G.~von Hippel, T.~Mendes, and R.~Sommer, {\it {O}n
  the generalized eigenvalue method for energies and matrix elements in lattice
  field theory},  {\em JHEP} {\bf 0904} (2009) 094,
  [\href{http://arxiv.org/abs/0902.1265}{{\tt arXiv:0902.1265}}].

\bibitem{Peardon:2009gh}
{\bf Hadron Spectrum Collaboration}, M.~Peardon, J.~Bulava,
  J.~Foley, C.~Morningstar, J.~Dudek, R.~G. Edwards, B.~Joo, H.-W. Lin, D.~G.
  Richards, and K.~J. Juge, {\it {A novel quark-field creation operator
  construction for hadronic physics in lattice QCD}},  {\em Phys. Rev. D} {\bf
  80} (2009) 054506, [\href{http://arxiv.org/abs/0905.2160}{{\tt
  arXiv:0905.2160}}].

\bibitem{Estabrooks:1977xe}
P.~Estabrooks {\em et~al.}, {\it {Study of $K \pi$ scattering using the reactions
  $K^\pm p \to K^\pm \pi^+ n$ and $K^\pm p \to K^\pm \pi^- \Delta^{++}$ at
  13-GeV/c}},  {\em Nucl. Phys.} {\bf B133} (1978) 490.

\bibitem{Aston:1987ir}
D.~Aston, N.~Awaji, T.~Bienz, F.~Bird, J.~D'Amore, {\em et~al.}, {\it {A study
  of $K^- pi^+$ scattering in the reaction $K^- p \to K^- \pi^+ n$ at
  11-GeV/c}},  {\em Nucl.Phys.} {\bf B296} (1988) 493.

\bibitem{Jongejans:1973pn}
B.~Jongejans, R.~van Meurs, A.~Tenner, H.~Voorthuis, P.~Heinen, {\em et~al.},
  {\it {Study of the I = 3/2 $K- \pi$- Elastic scattering from the reaction
  $K^- p \to K^- \pi^- p pi^+$ at 4.25-GeV/c incident $K^-$ momentum}},  {\em
  Nucl. Phys.} {\bf B67} (1973) 381.

\bibitem{Wilson:2013xxx}
D.~Wilson,  {\it  $K\pi$ scattering from lattice QCD}, presented at the
31st Int. Symp. on Lattice Field Theory - LATTICE 2013,
July 29 - August 3, 2013, Mainz, Germany.

\end{thebibliography}
\end{document}